\documentclass[twocolumn]{aa}
\usepackage{graphicx}
\usepackage{txfonts}
\newcommand{\feh}{\mathrm{[Fe/H]}} 
\begin{document} 
\title{A Homogenous Set of Globular Cluster Relative Distances and Reddenings 
\thanks{Based on observations with the Hubble Space Telescope}}

\author{ 
A. Recio-Blanco\inst{1} 
\and 
G. Piotto\inst{1} 
\and 
F. De Angeli\inst{1} 
\and
S. Cassisi\inst{2,3} 
\and 
M. Riello\inst{1,4} 
\and  
M. Salaris\inst{5} 
\and 
A. Pietrinferni\inst{2} 
\and 
M. Zoccali\inst{6} 
\and 
A. Aparicio\inst{3}} 
 
\offprints{A. Recio-Blanco. \\ New address: Observatoire de la Cote d'Azur
  (France). e-mail: arecio@obs-nice.fr} 
 
\institute { 
Dipartimento di Astronomia, Universit\`a di  Padova, Vicolo  dell'Osservatorio  2, I-35122     Padova, Italy\\ 
\email{recio,piotto,deangeli,riello@pd.astro.it } 
\and 
INAF, Osservatorio Astronomico di Collurania, Via M. Maggini, 64100, Teramo, Italy\\ 
\email{cassisi,adriano@te.astro.it} 
\and  
Instituto de Astrof\'\i sica de Canarias, Via La\'ctea s/n, 382002 La Laguna Tenerife, Spain\\ 
\email{aaj@ll.iac.es} 
\and  
ESO, Karl-Schwarschild-Str. 2, D-85748 Garching bei M$\ddot{\rm u}$nchen, Germany\\  
\and  
Astrophysics Research Institute, Liverpool John Moores University, Twelve Quays House, Birkenhead, CH41 1LD, UK {ms@astro.livjm.ac.uk}
\and   
Departamento de Astronomia, P. Universidad Catolica, Av. Vicuna Mackenna 4860, 7832-0436 Macul, Santiago - Chile\\   
\email{mzoccali@astro.puc.cl} 
} 
 
\date{Received ...; accepted  ...} 
 
\abstract{We present distance  modulus and reddening determinations
for  72 Galactic  globular clusters  from the  homogeneous photometric
database of  Piotto et al.\ (2002), calibrated to  the HST flight {\it
F439W}  and {\it F555W}  bands.  The distances have been determined by 
comparison with theoretical absolute magnitudes of the ZAHB. For  low and  
intermediate metallicity
clusters, we  have estimated the  apparent Zero Age  Horizontal Branch
(ZAHB) magnitude from the RR Lyrae level. For metal rich clusters,
the ZAHB magnitude was obtained from the fainter envelope of the red
HB.  Reddenings have been estimated by
comparison of the HST colour-magnitude diagrams (CMD) with ground
CMDs of low reddening template clusters.  
The homogeneity  of both  the photometric  data and  the adopted
methodological  approach  allowed us to obtain highly accurate
relative cluster distances and reddenings.  
Our results are also compared with recent compilations in the literature.

 }
 
\authorrunning{Recio-Blanco et al.} 
 
\titlerunning{Globular cluster distances with HST photometry} 
 
\maketitle 
 
\section{Introduction} 
 
Galactic  Globular Clusters (GGCs)  are extremely  useful astronomical
probes. Because they  are the oldest objects for  which we can estimate
the age, GGCs are commonly  used to infer relevant information on both
the Galaxy formation timescale and the early Universe. Moreover, they
constitute a well suited laboratory to study both the evolution 
of low-mass stars, and stellar dynamics.

Two key parameters needed in GGC studies are their distances and
reddenings.  As an example, the use of the absolute magnitude of
turnoff stars in the cluster colour-magnitude-diagram (CMD) to
determine the cluster age (see, e.g. Vandenberg, Stetson \&
Bolte~1996; Salaris \& Weiss~1998, and references therein) needs an
accurate distance estimate.  Also, the comparison between various
observed features of their CMDs (e.g., the absolute magnitude of the
luminosity function red giant branch bump, or the level of the tip of the
red giant branch) and the theoretical counterparts does require a
preliminar knowledge of both the cluster reddening and the distance.

A definitive assessment of both the absolute and relative GGC distance
scale is still lacking, mostly  due to the fact that the 'traditional'
Population~II  standard  candle,   e.g.  the  brightness  of  RR~Lyrae
pulsating stars, is not yet reliably calibrated 
empirically (e.g. Cacciari 2003). This is 
partially due to the  paucity  of  RR~Lyrae stars in  the  solar
neighbourhood, and the consequent large errors in RR~Lyrae parallax
determinations, even for the recent $Hipparcos$ data (Groenewegen \&
Salaris~1999), and also to the existence of significant systematic and random
uncertainties in other less direct methods applied to determine the 
RR~Lyrae intrinsic brightness (Bono 2003).

The advent  of $Hipparcos$ parallaxes  has on the other  hand improved
the accuracy  of the GGC  distance determination by the  subdwarf Main
Sequence (MS) fitting technique (e.g..  Carretta et al.~2000).  The problem
here is that this method can  be reliably applied only to a handful of
low  reddening  clusters, with  deep and well calibrated  high
accuracy MS photometry; in addition, current uncertainties
on the  metallicity scale of both clusters  and subdwarfs, and  on the
cluster  reddenings  may  still  cause sizable  uncertainties  on  the
distances derived by means of  this method (compare, e.g., the results
by Carretta et al.~2000 with Reid~1997, 1998).

In order to assess the accuracy and reliability of the various methods
used to infer  GGC distances, it is important  to compare the distance
measurements   obtained  with as many as possible different  and   
independent  distance
indicators,  such  as   the  aforementioned  empirical  MS
fitting,  the  RR  Lyrae   method,  and  the  fitting  of  theoretical
Horizontal Branch  (HB) models to their  observational counterpart.  This
kind of comparison is  relevant not  only  for  checking  the
consistency  between the  various  distance indicators,  but also  for
verifying  the  reliability  of   the  adopted,  if  any,  theoretical
scenario, as in the case of distances based on the fit to 
HB models.
On this respect, we note that a database of {\it relative} distances and
reddenings is of extreme importance: once we have accurate absolute
distances and reddenings for a set of GGCs, this database can be
easily used to obtain the absolute values for all the other clusters.
 
In the last decade, we have been working on a long-term project aimed
at carrying out a detailed quantitative analysis of the various
evolutionary sequences in the CMD of GGCs. Our main goals include the
derivation of an accurate GGC relative age scale (Rosenberg et al.
1999, Piotto et al. 2000), and a test of the accuracy of theoretical
models for low-mass metal-poor stars.  The main body of this
investigation has been performed by adopting an homogeneous and
self-consistent photometric dataset (available at
http://dipastro.pd.astro.it/globulars), based on both ground based
observations (Rosenberg et al~2000a, 2000b), and Hubble Space
Telescope data (the HST snapshot catalogue: Piotto et al.~2002).  This
large observational database has also been used to investigate the
level of agreement between theory and observations concerning
evolutionary timescales (Zoccali and Piotto 2000), the brightness and
size of the luminosity function Red Giant Branch (RGB) bump (Zoccali
et al.~\cite{z99}; Bono et al.~2001; Riello et al.~2003), the mixing
length parameter (Palmieri et al.~2002), the initial helium content
(Zoccali et al.~2000; Cassisi et al.~2003; Salaris et al.~2004), the
HB morphology (Piotto et al. 1999), the blue straggler stellar
population (Piotto et al.~2004).  The majority of these works needed
an as accurate as possible distance and reddening determination, and
in most cases we used a new set of distances and reddenings, based on
our photometrically homogeneous HST snapshot database. 
In this paper we present and thoroughly discuss how we obtained the
distances and reddenings adopted in the works above mentioned.

Distance estimates have been obtained from the fitting of theoretical
HB models to the observed counterpart in the CMD. We accurately
measured the observed HB luminosity level and, in turn, the distance
modulus, for about 40\% of the total GGC population, covering most of
the GGC metallicity range.  Our relative distances and reddenings are
more accurate than previous compilations, because {\it they are based
on a homogeneous photometric database, and have been derived by
applying consistently the same technique to all clusters}. Moreover,
the theoretical HB models we employed (Pietrinferni et al. 2004) have
been computed accounting for the most updated input physics.

The plan of the paper is as follows: in Section~2, we describe briefly
the photometric database and the theoretical models. Section 3
presents the actual measurements and the values of the distance
moduli.  We compare our distance estimates with relevant data
available in the literature in Section 4 and, finally, the main
conclusions are summarized in Section 5.

\section{The observational and theoretical databases} 
\label{s_obs} 
        
\subsection{The cluster database} 

\noindent 
The distance determinations presented here are based on the large
photometric data set from Piotto et al.\ (2002), observed with HST in
the {\it F439W} and {\it F555W} bands, calibrated to the WFPC2 flight
system.  The complete database includes a total of 74 GGCs, and
represents an unique opportunity to measure fundamental parameters of
GGCs.
 
The   observations,    preprocessing,   photometric   reduction,   and
calibration of  the instrumental magnitudes to the  HST flight system,
as well  as the artificial  star experiments performed to  derive the star
count completeness,  are reported in  full details in Piotto  et al.\
(2002).  For the  purpose of this paper, we  point out 
that  all  the  data  have been  processed  following  the  same
reduction steps: after the pre-processing, the instrumental photometry
for each cluster was obtained with DAOPHOT II/ALLFRAME (Stetson, 1987;
Stetson, 1994), the correction for  the CTE effect and the calibration
to the  flight system was accomplished following  the prescriptions by
Dolphin (2000).
 
\subsection{The GGCs metallicity scale} 

\noindent 
One of the pivotal problems in estimating both distances and ages for
GGCs is the adopted metallicity scale. As recently stated by Rutledge,
Hesser \& Stetson (1997, see also VandenBerg~\cite{v00}; Caputo \&
Cassisi \cite{cc02} and Kraft \& Ivans \cite{ki03}) current estimates
of the [Fe/H] values for GGCs are affected by an uncertainty of the
order of at least 0.15 dex. The situation becomes even more
uncertain when we consider the $\alpha-$element enhancement in
GGC stars: the measurements of $\alpha-$elements are affected by both
random and systematic uncertainties, they have been obtained in an
heterogeneous way, and only for a very limited number of GGCs.
 
In order to properly account for these unavoidable drawbacks we
decided, as in our previous works, to adopt the two most widely used
scales for the metal abundance in GGCs: the Zinn \& West (1984) scale
(hereinafter ZW), and the Carretta \& Gratton (1997, hereinafter CG)
one.  As for the $\alpha-$element enhancement, due to the lack of
self-consistent and accurate measurements for a sizeable sample of
GGCs, we adopt the following assumption: a mean [$\alpha$/Fe]=0.3 dex
for metal-poor and metal-intermediate clusters $(\feh<-1.0)$, and
[$\alpha$/Fe]=0.2 dex for metal-rich clusters $(\feh\ge-1.0)$.  The
choice of the former value is based on the estimates provided by
Carney (\cite{carney}), while the latter is obtained as a mean between
the values collected by Carney (\cite{carney}) and by Salaris \&
Cassisi (\cite{sc96}).
 
In order to estimate the global cluster metallicity by accounting for
the proper [Fe/H] value, and the chosen $\alpha-$element enhancement,
we have adopted the prescriptions provided by Salaris, Chieffi \&
Straniero~(\cite{scs93}), i.e.:
 
\begin{center} 
[M/H] $= [Fe/H] + log(0.638 f + 0.362)$ ; $\log{f}=[\alpha/Fe]$ 
\end{center} 
 
We assume an uncertainty of the order of $\pm$ 0.15 dex on [M/H], which
accounts for the uncertainties on both $\feh$ and
$[\alpha/\mathrm{Fe}]$ measurements (Rutledge et al. 1997).

\subsection{The theoretical framework} 
 
The theoretical predictions adopted in this investigation are based on
the updated set of stellar models by Pietrinferni et al.~(2004), and
we refer the interested reader to that paper for a complete
discussion about these models\footnote{All the theoretical models
adopted in present work as well as a more extended set of evolutionary
results and isochrones can be found at the URL site:
http://www.te.astro.it/BASTI/index.php.}.  For the purposes of this
paper, we briefly list the main changes in the adopted physical inputs
with respect to previous works (Cassisi \& Salaris~1997):
 
\begin{itemize} 
\item 
the radiative opacity is obtained from the OPAL tables (Iglesias \&
Rogers~1996) for temperatures larger than $10^4$ K, and from Alexander
\& Ferguson (\cite{alexander}) for lower temperatures.  Conductive
opacity for electron degenerate matter is computed following Potekhin
(\cite{pot}).
 
\item 
We updated the energy loss rates for plasma-neutrino processes by
using the most recent and accurate results provided by Haft, Raffelt
\& Weiss~(\cite{haft}).  For all other processes we still rely on the
same prescriptions adopted by Cassisi \& Salaris (1997).
 
\item 
The nuclear reaction rates have been updated by using the NACRE
database (Angulo et al.~1999), with the exception of the
$^{12}$C$(\alpha,\gamma)^{16}$O reaction. For this reaction we now
adopt the more accurate recent determination by Kunz et
al.~(\cite{kunz}).
 
\item 
The accurate Equation of State (EOS) by A. Irwin has been used.  An
exhaustive description of this EOS is still in preparation (Irwin et
al.~2004) but a brief discussion of its main characteristics can be
found in Cassisi, Salaris \& Irwin (\cite{csi03}). It is enough to
mention here that this EOS, whose accuracy and reliability is similar
to the OPAL EOS developed at the Livermore Laboratories (Rogers,
Swenson \& Iglesias \cite{rsi96}) and recently updated in the
treatment of some physical inputs (Rogers \& Nayfonov \cite{rn02}),
allows us to compute self-consistent stellar models in all
evolutionary phases relevant to present investigation.
 
\item 
The extension of the convective zones is fixed by means of the
classical Schwarzschild criterion.  Induced overshooting and
semiconvection during the He-central burning phase are accounted for
following Castellani et al.~(\cite{cast85}). The thermal gradient in
the superadiabatic regions is determined according to the mixing
length theory, whose free parameter has been calibrated by computing a
solar standard model.
 
\item 
The set of evolutionary models has been computed for metallicities in
the range: $0.0001\le{Z}\le0.04$. However, in the present work only the
models for metallicity equal or lower than the solar one have been
used.  We adopt the scaled-solar heavy element mixture (Grevesse \&
Noels~\cite{gn93}).

\item 
As far as the initial He-abundance is concerned, we adopt the estimate
recently provided by Salaris et al.~(\cite{s04}) on the basis of
new measurements of the $R$ parameter in a large sample of
GGCs\footnote{We recall that this cluster database is exactly the same
adopted in the present work.}. They found an initial He-abundance for
GGC stars of the order of $Y=0.245$, which is in fair agreement with
recent empirical measurements of the cosmological baryon density
provided by W-MAP (Spergel et al.~2003). To reproduce the calibrated
initial solar He-abundance we used $dY/dZ\approx1.4$ (Pietrinferni et
al.~2004).
 
\item  
For each fixed chemical composition, we have adopted the He core mass
and the surface He abundance at the RGB tip of a star igniting central
He burning at an age of about $\sim12$ Gyr.  Once the RGB progenitor
mass is chosen, a suitable set of Zero Age Horizontal Branch (ZAHB)
models for different assumptions about the mass of the H-rich stellar
envelope -- i.e., about the efficiency of mass loss during the RGB
phase -- has been computed. In brief, the initial models of our HB
sequences have a fully homogeneous H-rich envelope around the He core
mass of the selected progenitor; the proper ZAHB model is obtained
when all the secondary elements involved in the H-burning through the
CNO-cycle are relaxed to their equilibrium values.
 
\item 
Bolometric magnitudes have been transformed to HST F555W magnitudes
according to the transformations provided by Origlia \& Leitherer
(\cite{origlia}), which based on the atmosphere models computed by
Bessell, Castelli \& Plez (\cite{bcp98}).

\end{itemize} 
 
\noindent 
From  the computed  ZAHB models,   we have   estimated  the ZAHB
brightness  at the level  of the RR Lyrae  instability strip, i.e., at
$\log{T_e}\approx 3.85$. In Table~1 we list, for each assumed chemical
composition, the bolometric magnitude, the star mass and the HST F555W
magnitude of the ZAHB at $\log{T_e}= 3.85$.
 
By performing a quadratic regression to these data,  we obtain 
the following dependence of the ZAHB F555W magnitude on the  stellar
total metallicity:

\begin{center} 
 M$^{ZAHB}_{F555W} = 0.981  + 0.410 [M/H] + 0.061 [M/H]^2$ ~~~~~~{1)}
\end{center}  
with   $r^2=0.99$,   which  is   valid   in   the  metallicity   range:
$-2.3\le[M/H]\le0.06$\footnote{We notice  that  the
'solar  metallicity' models correspond to [M/H]=0.06, instead  of 0.0.  The
reason is  that our adopted  models do not include diffusion - which we
know is active in the Sun, but according to some empirical evidence
(Bonifacio et al.~2002 and references therein) is possibly inhibited  
at least at the surface  of low-mass, metal-poor stars. When  diffusion  
is included  the  'solar  metallicity'
composition  would  provide  [M/H]=0.0  only  at  the  solar  age  for
solar-like models.}.
 
The models adopted in this work have been computed by neglecting
atomic diffusion. However, Castellani et al.\ (1997) and Cassisi et
al.\ (1998) have shown that the effect of atomic diffusion on the ZAHB
brightness at the level of the RR Lyrae instability strip is to
decrease it of about $\Delta{\log}(L/L_\odot)\approx0.02$, i.e.,
$\Delta{F555W}\approx+0.05$ mag.  This means that, if we would account
for the occurrence of atomic diffusion, the derived distance modulus
estimates (see next section) would be decreased only by about 0.05
mag.

As stated above, our theoretical ZAHB luminosities are based on
updated stellar models, computed by accounting for the \lq{best}\rq\
physics presently available. However we are aware of remaining
uncertainties affecting the prediction of the HB brightness, as
discussed by several authors (Vandenberg et al. 2000, De Santis \&
Cassisi 1999, Cassisi et al. 1998 and references therein). We refer
the reader to the quoted papers for deeper analysis on this
subject. Here we will compare the distances obtained using our ZAHB
models with independent empirical determinations, in order to test the
accuracy of the models we employed.
  
%%%%%%%%%%%%%%%%%%%%%%%%%%%%%%%%%%%%%%%%%%%%%%%%%%%%%%%%%%%%%%%%%%%%%%%% 
\begin{table}[!t] 
\begin{center} 
\caption{Theoretical predictions for the ZAHB luminosity, evolutionary
mass and absolute F555W magnitude at the RR Lyrae instability strip as
a function of $[M/H]$.}
\label{tabzahb} 
\begin{tabular}{cccc} 
\hline 
\hline 
\noalign{\smallskip} 
$[M/H]$ & $M/M_\odot$ & $\log(L/L_\odot)$ & F555W (mag) \\ 
\noalign{\smallskip} 
\hline 
-2.27 & 0.821  &  1.780  &  0.351  \\ 
-1.79 & 0.721  &  1.732  &  0.468  \\ 
-1.27 & 0.650  &  1.687  &  0.564  \\ 
-0.96 & 0.620  &  1.653  &  0.634  \\ 
-0.66 & 0.594  &  1.614  &  0.721  \\ 
-0.25 & 0.565  &  1.540  &  0.884  \\ 
 0.06 & 0.543  &  1.489  &  0.988  \\ 
\hline 
\hline 
\end{tabular} 
\end{center} 
\end{table}  
%%%%%%%%%%%%%%%%%%%%%%%%%%%%%%%%%%%%%%%%%%%%%%%%%%%%%%%%%%%%%%%%%%%%%%%% 

\section{Distance moduli determination} 
\label{cc} 
 
The distance modulus for each cluster in the database was obtained by
comparing the F555W apparent ZAHB magnitude with the theoretical
absolute magnitude as obtained from equation 1).  A standard
method for deriving the ZAHB magnitude for both intermediate and
metal-poor clusters is to adopt the mean magnitude of the
corresponding RR Lyrae stars. This however was not possible in our
case because the Piotto et al.~(2002) photometry covers a very short
time interval and, as a consequence, the RR Lyrae stars were always
measured at random pulsation phases.
 
In order to overcome this problem, we have undertaken an approach
similar to the one already used by Zoccali et al.\ (1999)\footnote{ We 
refer the reader to the quoted reference for a detailed discussion on 
the difficulty of measuring the ZAHB luminosity at the level of the 
RR Lyrae instability strip in those clusters characterized
by a very blue or red horizontal branch.}. 
We first
divided the clusters into two samples on the basis of their
metallicity: the low- and intermediate-metallicity (hereinafter LIM)
clusters ($[Fe/H]<-1.0$), and the metal-rich (hereinafter MR) clusters
($[Fe/H]\ge-1.0$).  In the following, we describe in details the
different approaches we used to estimate the ZAHB level for the two
different cluster samples.

\begin{figure} 
\centerline{\includegraphics[width= 8 cm]{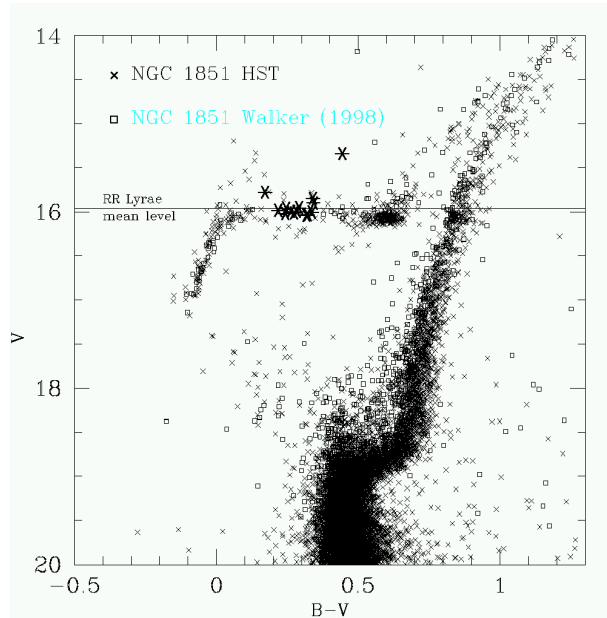}}
\protect\caption[] {Comparison between the ground-based CMD of NGC1851 
obtained by Walker (1998) and the HST CMD one, transformed to the standard
Johnson system. The location of the RR Lyrae stars as observed by Walker (1998) is
also shown (asterisks). The horizontal line displays the mean RR Lyrae luminosity level.}
\end{figure} 
 
\begin{figure} 
\centerline{\includegraphics[width= 8cm]{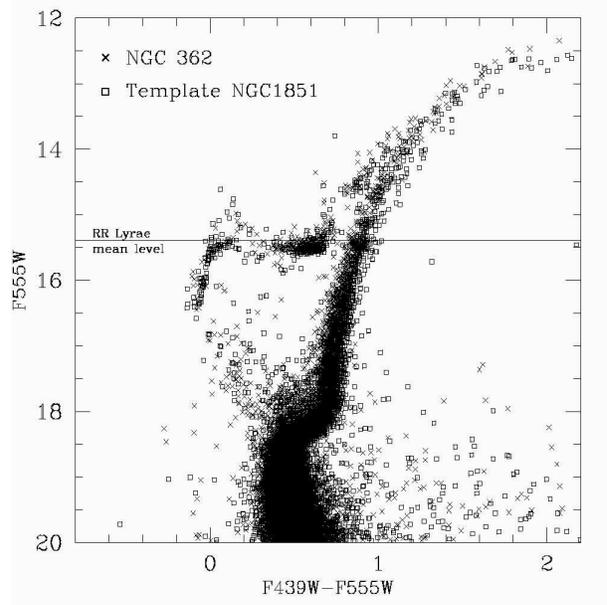}} 
\protect\caption[] 
{Determination of the mean RR Lyrae F555W magnitude in the cluster NGC362 
by comparing its CMD with the template CMD of NGC1851. The horizontal line
displays the estimated mean RR Lyrae magnitude.} 
\end{figure} 
 
\subsection{The ZAHB luminosity level for LIM clusters}

Since the HB morphology does strongly depend on the cluster metallicity, we
selected  five   clusters,  all  with  metallicity  lower   than  -- or
approximately   equal  to  --  [Fe/H]=$-1.0$,   to  use   as  template
clusters.  They   have  been  selected  according   to  the  following
prescriptions:

\begin{itemize}

\item{low interstellar reddening;}

\item{a sizeable population of RR Lyrae variables;}

\item{accurate ground-based photometric  data for  both  static and
pulsating stars.}

\end{itemize}

\noindent
The selected  clusters are NGC~1851( Walker,  1998), NGC~4590 (Walker
1994), NGC~5272 (Buonanno et al.  1994), NGC~5904 (Caputo et al. 1999)
and NGC~6362 (Walker 2001, priv. com.).

By using the histogram of the observed RR Lyrae mean magnitudes, we
estimated the mean RR Lyrae luminosity level in the standard Johnson
system, for all the five clusters selected from the literature.  These
clusters will be used to determine the ZAHB level in other GGCs,
that cannot fulfill all three conditions listed before, within a
narrow metallicity range around the template ones.

The metallicity of the templates on the ZW scale,
the [Fe/H] range within which they have been employed, and the
mean $V$ and $F555W$ magnitudes of their RR~Lyrae stars 
are listed in Table \ref{templates}.

We took care  that all the selected template  clusters had a reddening
E($B-V$)$<$0.1,  in order  to  minimize  calibration  errors in  the
determination  of the  RR~Lyrae level  when comparing  the ground-based
CMDs with the HST snapshot ones transferred in the Johnson system (see
below).

%The method for estimating the mean RR Lyrae luminosity level in the F555W HST band, 
%adopted for all LIM clusters, is the following:
The method for estimating the ZAHB luminosity level in the F555W HST band, 
adopted for all LIM clusters, is the following:

\begin{itemize}

\item{the mean RR Lyrae luminosity level in the ground-based Johnson
system for the template clusters has been translated into the HST {\sl
flight} photometric system. Due to the fact that there are
non-negligible differences between the standard ground-based Johnson
photometry and the HST {\sl flight} photometry, this has been
accomplished following a two-step procedure. 
As a first step, 
we have superposed the ground-based CMD to the corresponding HST
snapshot CMD calibrated to the Johnson system. This has allowed us
to set the RR~Lyrae mean magnitude measured on the groundbased CMD 
on the HST snapshot CMD. After this, we have transferred the RR~Lyrae
mean level to the CMD in the WFPC2 flight system. This allowed us to
measure the RR~Lyreae mean $F555W$ magnitude in the WFPC2 flight
system.
The mean apparent magnitude of the template cluster RR Lyrae stars 
has been then transformed into the apparent ZAHB  magnitude by accounting
for the formula by  Cassisi \& Salaris (1996)\footnote{It is important
to notice  that this relation  was originally obtained for  the V-band
magnitude.  However,  by using  HB models and  synthetic CMDs  we have
verified that it is valid also for the F555W band}.
The use of this relation is particularly justified by the fact that 
all the template clusters have a sizeable population of RR Lyraes stars.
\begin{center} 
 m$^{ZAHB}_{F555W} = m^{RR-Lyrae}_{F555W} + 0.152 + 0.041 [M/H]$ 
\end{center} 
NGC~5272, was not
included in the Piotto et al.~(2002) database. For that reason, we overlapped 
its CMD
to the snapshot CMD of NGC~1904 ([Fe/H]=$-1.6$) in order to fix the RR
Lyrae level, and then adopted the snapshot CMD of NGC~1904 as
template to determine the ZAHB level of the clusters in the
corresponding (see table.~\ref{templates}) metallicity interval.}

%\item{the mean RR Lyrae $F555W$ magnitude for each remaining
\item{the ZAHB $F555W$ magnitude for each remaining
cluster in our sample has been determined in the following way.  The
appropriate template HST snapshot CMD calibrated to the flight system
has been shifted in both magnitude and colour over the CMD of the
cluster, until their HBs overlap, as illustrated in Fig.~2.  In
particular, we have been careful to match the region of the blue HBs
around the point where the HB becomes vertical in the ($F555W$,
$F439W-F555W$) plane. The location of this point on the CMD depends
only on the bolometric correction, and is independent on metallicity
(Brocato et al.\, 1998).  This procedure allowed us to obtain both the
m$^{ZAHB}_{F555W}$ magnitude and the E($F439W-F555W$) reddening
relative to the template one.}

\end{itemize}

\begin{table*}[!t] 
\begin{center} 
\caption{Cluster templates for the measurement of the average RR Lyrae
level. The estimated error in the mean RR Lyrae F555W magnitude is of the order of
0.05 mag.}
\label{templates} 
\begin{tabular}{ccccc} 
\hline 
\hline 
\noalign{\smallskip} 
Cluster & [Fe/H] & metallicity range & $<V(RR)>$ & $<F555W(RR)>$\\ 
\noalign{\smallskip} 
\hline 
NGC~6362 & $-1.1$ & $-$1.1$<$[Fe/H]$<-$0.8    & 15.29 & 15.30 \\
NGC~1851 & $-1.2$ & $-$1.3$<$[Fe/H]$\leq-$1.1 & 16.04 & 16.04 \\
NGC~5904 & $-1.4$ & $-$1.5$<$[Fe/H]$\leq-$1.3 & 15.07 & 15.10 \\
NGC~1904 & $-1.7$ & $-$1.8$<$[Fe/H]$\leq-$1.5 & 16.17 & 16.16 \\    
NGC~4590 & $-2.1$ &      [Fe/H]$\leq -1.8$    & 15.58 & 15.65 \\ 
\hline 
\hline 
\end{tabular} 
\end{center} 
\end{table*} 

\begin{figure} 
\centerline{\includegraphics[width= 8 cm]{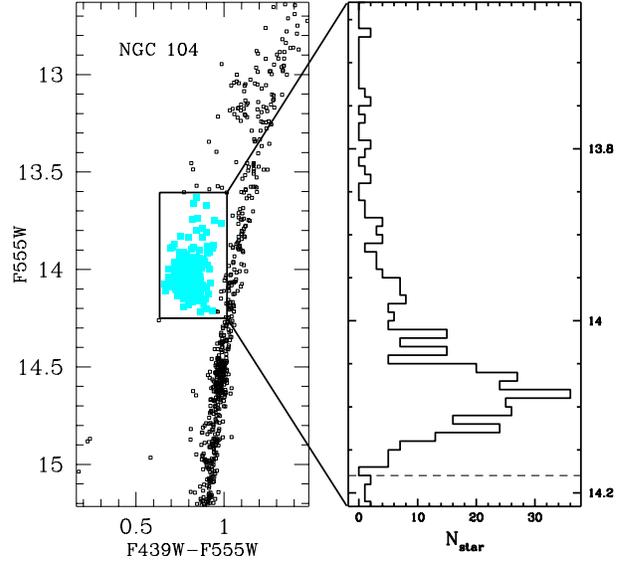}} 
\protect\caption[] 
{Determination of F555W magnitude of the ZAHB for the metal-rich
cluster NGC~104. {\sl Left panel}: the HST CMD of the cluster. {\sl Right panel}: the
histogram of the stellar cluster population within the box shown in the left panel.} 
\end{figure} 

\subsection{High metallicity clusters}
 
For clusters with [Fe/H] $\ge$ -1.0, showing generally a red HB and no
RR Lyrae stars, the magnitude m$^{ZAHB}_{F555W}$ was derived 
according to the following relation (see Zoccali
et al.\ (1999) for more details):
\begin{center} 
  m$^{ZAHB}_{F555W} = m^{le}_{F555W}  - 3\sigma_{F555W}$ 
\end{center} 
where  m$^{le}_{F555W}$  is  the  magnitude  of  the  lower  (fainter)
envelope of the red HB,  previously determined on the histogram of the
magnitude   distribution    of   the    HB   stars, as shown in Figure 3, and 
$\sigma_{F555W}$  is the  photometric error at  the HB magnitude interval,
estimated  through artificial star experiments. In general,
$\sigma_{F555W}$ is of  the order  of  0.02  mag, representing  very  
small deviations  in magnitude.
Where possible (i.e. for clusters with a red HB, but a metallicity low
enough to be  able to determine the RR Lyrae  level by comparison with
the template  clusters), we  have verified that  the two  methods give
consistent ZAHB magnitudes.
 
The final apparent distance moduli for all our clusters were
calculated by simply computing the difference (m$^{ZAHB}_{F555W}$ -
M$^{ZAHB}_{F555W}$), where the absolute ZAHB magnitudes were obtained
from Eq.~1.  Tables~3 and 4 list the distance moduli and reddenings
obtained for our data set. Note that, due to the CMD photometry accuracy,
we could measure the distance to 72 out of the 74 clusters in the snapshot sample. 
For 2 clusters, field star contamination and/or differential reddening
made impossible to measure the ZAHB magnitude.

\begin{table*}[!t]
%\footnotesize{\begin{tabular}{ccccccc} 
\begin{center} 
\begin{tabular}{ccccccc}
\hline
\hline  
  ID  &  m(ZAHB) &  (m-M)    & \tiny{Reddening} & E(B-V) & [Fe/H] & [Fe/H] \\                 
      &  $F555W$ &  $F555W$  &                  &        &   ZW   &   CG   \\      
\hline 
   ic1257  &  20.20     &   19.69$\pm$0.10   &  0.71   & 0.72  & ....... &   ...... \\
   ic4499  &  17.75     &   17.22$\pm$0.10   &  0.15   & 0.15  & -1.5    &    -1.29 \\
   n0104   &  14.16     &   13.41$\pm$0.08   &  .......&.......& -0.71   &    -0.70 \\
   n0362   &  15.50     &   14.85$\pm$0.10   &  0.005  & 0.004 & -1.27   &    -1.15 \\
   n1261   &  16.85     &   16.26$\pm$0.09   &  0.01   & 0.01  & -1.31   &    -1.17 \\
   n1851   &  16.15     &   15.52$\pm$0.10   &  0.02   & 0.02  & -1.36   &    -1.20 \\
   n1904   &  16.25     &   15.71$\pm$0.08   &  0.01   & 0.01  & -1.69   &    -1.37 \\
   n2808   &  16.35     &   15.70$\pm$0.09   &  0.14   & 0.14  & -1.37   &    -1.21 \\
   n3201   &  14.85     &   14.32$\pm$0.08   &  0.24   & 0.25  & -1.61   &    -1.23 \\
   n4590   &  15.73     &   15.30$\pm$0.08   &  0.05   & 0.05  & -2.09   &    -1.99 \\
   n4147   &  16.98     &   16.50$\pm$0.08   &  0.01   & 0.01  & -1.80   &    -1.55 \\
   n4372   &  15.58     &   15.15$\pm$0.09   &  0.40   & 0.42  & -2.08   &    -1.93 \\
   n4833   &  15.70     &   15.21$\pm$0.08   &  0.28   & 0.29  & -1.86   &    -1.58 \\
   n5024   &  16.83     &   16.38$\pm$0.08   &  0.01   & 0.01  & -2.04   &    -1.86 \\
   n5634   &  17.68     &   17.20$\pm$0.09   &  0.01   & 0.01  & -1.82   &    -1.57 \\
   n5694   &  18.53     &   18.06$\pm$0.08   &  0.12   & 0.12  & -1.92   &    -1.69 \\
   n5824   &  18.53     &   18.06$\pm$0.08   &  0.12   & 0.12  & -1.87   &    -1.63 \\
   n5904   &  15.20     &   14.59$\pm$0.08   &  0.03   & 0.03  & -1.40   &    -1.11 \\
   n5927   &  16.79     &   15.90$\pm$0.08   &  .......&.......& -0.30   &    -0.14 \\
   n5946   &  17.60     &   17.01$\pm$0.09   &  0.63   & 0.64  & -1.37   &    -1.21 \\
   n5986   &  16.70     &   16.17$\pm$0.08   &  0.26   & 0.27  & -1.67   &    -1.42 \\
   n6093   &  16.35     &   15.85$\pm$0.08   &  0.21   & 0.22  & -1.68   &    -1.43 \\
   n6139   &  18.05     &   17.54$\pm$0.09   &  0.71   & 0.72  & -1.65   &    -1.40 \\
   n6171   &  15.79     &   15.11$\pm$0.09   &  0.48   & 0.49  & -0.99   &    -0.97 \\
   n6205   &  15.05     &   14.51$\pm$0.08   &  0.01   & 0.01  & -1.65   &    -1.39 \\
   n6218   &  14.80     &   14.24$\pm$0.20   &  0.21   & 0.22  & -1.61   &    -1.37 \\
   n6229   &  18.10     &   17.53$\pm$0.09   &  0.02   & 0.02  & -1.54   &    -1.32 \\
   n6235   &  17.00     &   16.42$\pm$0.09   &  0.30   & 0.31  & -1.40   &    -1.22 \\
   n6256   &  18.22     &   17.42$\pm$0.09   &  1.18   & 1.18  & ....... &    ..... \\
   n6266   &  16.30     &   15.69$\pm$0.09   &  0.53   & 0.54  & -1.28   &    -1.15 \\
   n6273   &  16.55     &   16.04$\pm$0.08   &  0.33   & 0.34  & -1.68   &    -1.43 \\
   n6284   &  17.50     &   16.90$\pm$0.08   &  0.28   & 0.29  & -1.40   &    -1.22 \\
   n6287   &  17.13     &   16.70$\pm$0.08   &  0.65   & 0.66  & -2.05   &    -1.88 \\
   n6293   &  16.48     &   16.02$\pm$0.08   &  0.61   & 0.62  & -1.92   &    -1.69 \\
   n6304   &  16.39     &   15.58$\pm$0.09   &  .......&.......& -0.59   &    -0.60 \\
   n6316   &  17.92     &   17.07$\pm$0.10   &  .......&.......& -0.47   &    -0.44 \\
   n6325   &  18.05     &   17.41$\pm$0.09   &  1.02   & 1.03  & -1.44   &    -1.25 \\
\hline
%\end{tabular}}
\end{tabular}
\caption{Apparent ZAHB F555W magnitude (Col. 2), corresponding 
distance modulus in F555W (Col. 3), the reddening in flight system E(F439W-F555W) (Col. 4), the 
reddening in Johnson system E(B-V) (Col. 5) and the
metallicity in the Zinn \& West and Carretta \& Gratton scales (Col. 6 and 7
respectively). }  
\end{center} 
\end{table*} 

\begin{table*}[!t]
%\footnotesize{\begin{tabular}{ccccccc} 
\begin{center}
\begin{tabular}{ccccccc}
\hline
\hline   
 ID & m(ZAHB) &  (m-M)  &\tiny{Reddening}& E(B-V) & [Fe/H] & [Fe/H] \\                
    & $F555W$ & $F555W$ &                &        &   ZW   &   CG   \\ 
\hline           
 n6342   &  17.07      &   16.28$\pm$0.09   &  .......&.......& -0.62   &    -0.64\\    
 n6355   &  17.80      &   17.25$\pm$0.10   &  0.77   & 0.78  & -1.5    &    -1.29\\
 n6356   &  17.65      &   16.81$\pm$0.09   &  .......&.......& -0.62   &    -0.64\\
 n6362   &  15.42      &   14.71$\pm$0.09   &  0.08   & 0.08  & -1.08   &    -0.96\\
 n6380   &  19.62      &   18.75$\pm$0.15   &  1.58   & 1.58  & -1.00   &    -0.98\\
 n6388   &  17.99      &   16.49$\pm$0.09   &  .......&.......& -0.74   &    -0.77\\
 n6401   &  17.85      &   17.19$\pm$0.15   &  0.90   & 0.91  & -1.13   &    -1.06\\
 n6402   &  17.40      &   16.82$\pm$0.09   &  0.65   & 0.66  & -1.39   &    -1.22\\
 n6440   &  18.92      &   17.99$\pm$0.09   &  .......&.......& -0.26   &    -0.06\\
 n6441   &  17.96      &   17.13$\pm$0.09   &  .......&.......& -0.59   &    -0.60\\
 n6453   &  17.80      &   17.25$\pm$0.09   &  0.61   & 0.62  & -1.53   &    -1.31\\
 n6517   &  19.20      &   18.61$\pm$0.09   &  1.23   & 1.23  & -1.34   &    -1.19\\
 n6522   &  16.80      &   16.23$\pm$0.09   &  0.53   & 0.54  & -1.44   &    -1.25\\
 n6539   &  18.52      &   17.71$\pm$0.09   &  .......&.......& -0.66   &    -0.69\\
 n6540   &  15.95      &   15.31$\pm$0.09   &  0.52   & 0.53  & ....... &  .......\\
 n6544   &  15.25      &   14.71$\pm$0.08   &  0.76   & 0.77  & -1.56   &    -1.33\\
 n6569   &  17.62      &   16.91$\pm$0.08   &  0.56   & 0.57  & -0.86   &    -0.88\\
 n6584   &  16.60      &   16.04$\pm$0.09   &  0.02   & 0.02  & -1.54   &    -1.32\\ 
 n6624   &  16.11      &   15.24$\pm$0.10   &  .......&.......& -0.35   &    -0.23\\
 n6637   &  16.04      &   15.27$\pm$0.09   &  .......&.......& -0.59   &    -0.60\\
 n6638   &  16.97      &   16.27$\pm$0.10   &  0.38   & 0.39  & -1.15   &    -1.08\\
 n6642   &  16.70      &   16.11$\pm$0.09   &  0.43   & 0.44  & -1.29   &    -1.16\\
 n6652   &  16.06      &   15.38$\pm$0.08   &  0.13   & 0.13  & -0.89   &    -0.90\\
 n6681   &  15.75      &   15.20$\pm$0.08   &  0.06   & 0.06  & -1.51   &    -1.30\\
 n6712   &  16.22      &   15.53$\pm$0.09   &  0.38   & 0.39  & ....... &  .......\\
 n6717   &  15.80      &   15.19$\pm$0.20   &  0.22   & 0.23  & -1.32   &    -1.18\\
 n6723   &  15.55      &   14.89$\pm$0.10   &  0.05   & 0.05  & -1.09   &    -1.04\\
 n6760   &  17.92      &   17.09$\pm$0.08   &  .......&.......& -0.52   &    -0.51\\
 n6838   &  14.54      &   13.75$\pm$0.10   &  .......&.......& -0.58   &    -0.70\\
 n6864   &  17.70      &   17.10$\pm$0.09   &  0.20   & 0.21  & -1.32   &    -1.18\\
 n6934   &  16.95      &   16.41$\pm$0.10   &  0.08   & 0.08  & -1.54   &    -1.32\\
 n6981   &  16.90      &   16.32$\pm$0.08   &  0.05   & 0.05  & -1.54   &    -1.32\\
 n7078   &  15.88      &   15.48$\pm$0.11   &  0.09   & 0.09  & -2.15   &    -2.12\\
 n7089   &  16.03      &   15.50$\pm$0.09   &  0.01   & 0.01  & -1.62   &    -1.38\\
 n7099   &  15.23      &   14.81$\pm$0.08   &  0.03   & 0.03  & -2.13   &    -1.91\\
\hline
%\end{tabular}}
\end{tabular}
\caption{Apparent ZAHB F555W magnitude (Col. 2), corresponding 
distance modulus in F555W (Col. 3), the reddening in flight system E(F439W-F555W) (Col. 4), the 
reddening in Johnson system E(B-V) (Col. 5) and the
metallicity in the Zinn \& West and Carretta \& Gratton scales (Col. 6 and 7
respectively). }  
\end{center} 
\end{table*} 

For each cluster, the total E($F439W-F555W$) reddening was determined
by adding the shift in colour of the CMD fit to the reddening of the
corresponding template. For the metal rich clusters ([Fe/H]$>$-0.8), no reddening determination
was therefore possible. 

The E($F439W-F555W$) reddening of the templates was calculated
starting from the E($B-V$) values tabulated by Harris~(2003), and by
interpolating the relationship given by Holtzman et al.\ (1995)
between E($B-V$) and the extinctions A$_{F555W}$ and A$_{F439W}$.

The error  in (m-M)$_{F555W}$ has  been determined considering:  i) the
error  in  matching  the CMDs of the templates to find the RR Lyrae
level in the flight system ($\sim$  0.04  magnitudes), ii) the error in
matching the CMD of the template with that of the object cluster
($\sim$  0.07  magnitudes),  iii)  the
photometric error  at the  level of the  ZAHB ($\sim$ 0.02 magnitudes),  iv) the
standard deviation of the mean of the RR~Lyrae magnitudes, and v) the
photometric error of the template clusters. The first two errors where
determined repeating several times the match and 
considering the scatter in the results.

% Francesca adds the following sentence
Column 5 of Tables~3 and 4 lists our reddening values transformed to 
the Johnson system by interpolating the relationship given by Holtzman et al.\ (1995)
between the extinctions A$_{F555W}$ and A$_{F439W}$ and E($B-V$).

\begin{figure*} 
\centerline{\includegraphics[width=11cm]{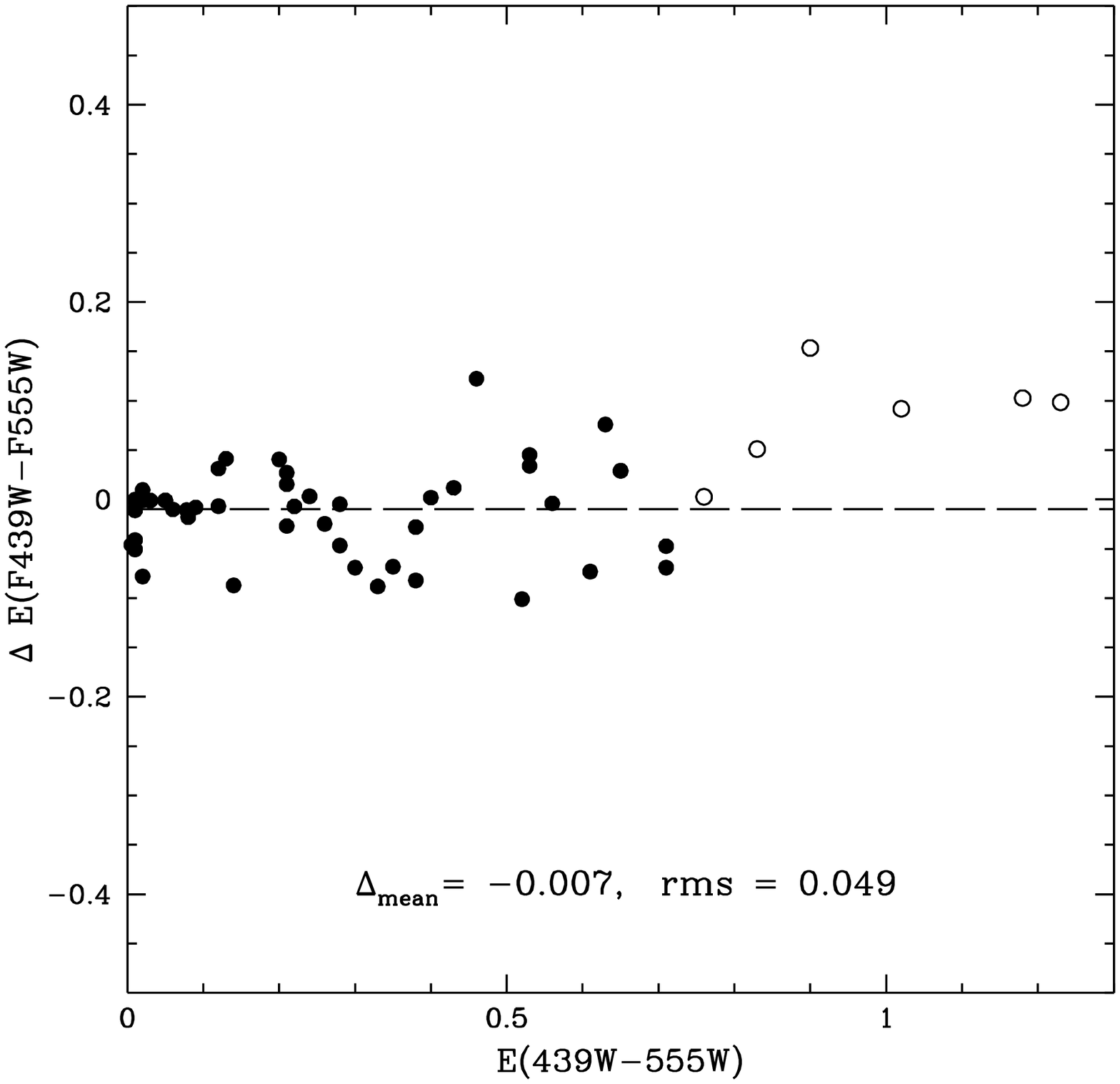}}
\protect\caption[] {Comparison of the reddenings from this paper with
those of Harris~(2003).  The E($B-V$) values of Harris~(2003) have been
transformed into the corresponding E($F439W-F555W$) reddenings by
calculating the extinctions coefficents A$_{F555W}$ and A$_{F439W}$
following Holtzman et al.\ (1995).  The differences are plotted as a
function of the cluster reddening. Clusters with  E($F439W-F555W$)$>$0.75 
(open symbols) are not considered in the mean.}
\end{figure*} 

\begin{figure*} 
\centerline{\includegraphics[width=11cm]{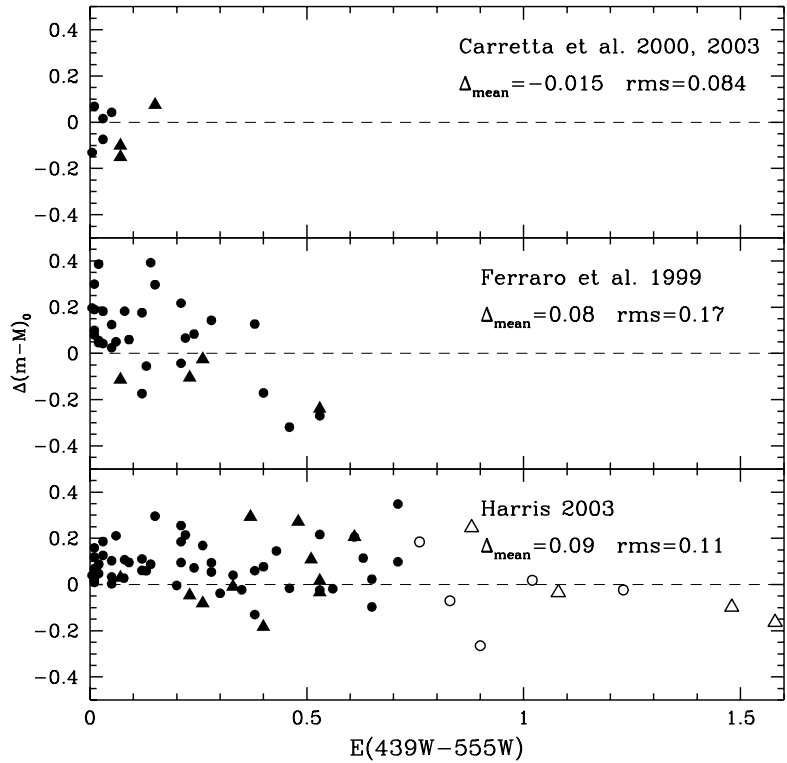}}
\protect\caption[] {Comparison  of the  distance moduli from this paper
with  those of  Harris~(2003),  Ferraro et  al.\  (1999), Carretta  et
al.~(2000) and Carretta et al.\ (2003, filled triangles).  The
differences (our measurements with repect to the literature ones) are plotted as a function of the cluster
reddening. Clusters with E($F439W-F555W$)$>$0.75 (open symbols) or [Fe/H]$>$-0.8 (triangles)
have not been used in the calculation of the mean
differences and standard deviations.}
\end{figure*} 

\begin{figure*} 
\centerline{\includegraphics[width=14cm]{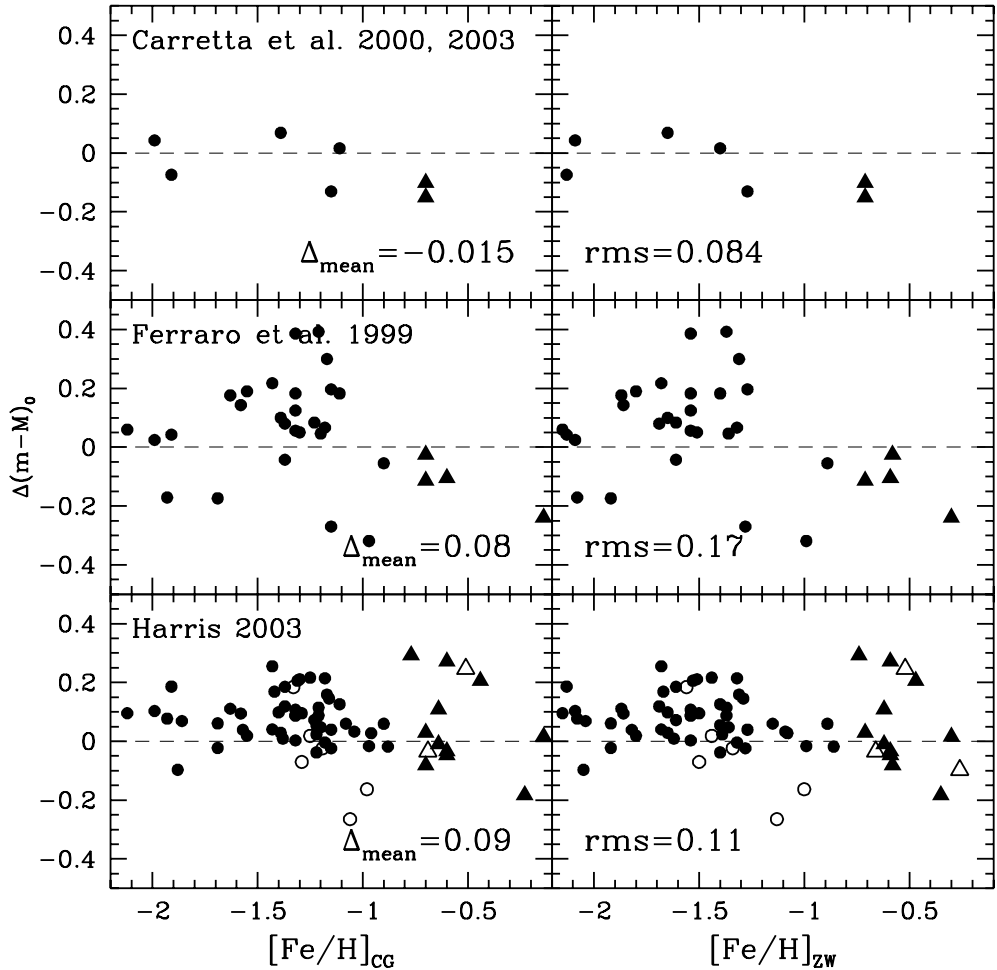}}
\protect\caption[] {Same as Figure 4, but with the differences plotted
as a function of metallicity from Carretta \& Gratton (1997) (left
column) panel, and Zinn \& West (1984) (right column). 
Clusters with
E($F439W-F555W$)$>$0.75 (open symbols) or [Fe/H]$>$-0.8 (triangles) have not been
used in the calculation of the mean differences and standard
deviations.}
\end{figure*} 

\section{Comparison with other datasets} 

The relative distances inferred from Tabs.~3 and 4 are
the most accurate ones that can be obtained for such a huge set
of GGCs with the present observational techniques\footnote{For a
limited number of clusters it is surely possible to have more accurate
relative and absolute distances, as shown by Carretta et al. (2000).}.
They are based on a photometrically homogeneous database, and they
have been obtained following the same methodology.  In this respect, the
empirical measurement of the apparent magnitude of the ZAHBs are
robust.  The assumptions we made in estimating our distance moduli are
clearly stated.  One can easily adopt a different relationship between
the theoretical ZAHB brightness and the metallicity, and obtain
apparent distances starting from the observed ZAHB levels listed in Tabs.~3 and 4. 
The same is valid for the reddenings, and consequently absolute distances.

In view of the 
significant 
importance of the GGC distance modulus
estimates, it is worth to compare the present results with others
in the astronomical literature.  We have selected three recent and
independent works dealing with GGC distances 
and reddenings: Harris~(2003), Ferraro
et al.~(1999), and Carretta et al.~(2000, 2003).  The Harris~(2003)
catalogue is a database of parameters for GGCs collected from the
literature, and therefore coming from photometrically inhomogeneous
CMDs, and based on different methods for the HB level measurements.
Although it represents a very useful tool for analizing the general
properties of the GGC system, this limitation has to be taken into
account.  Ferraro et al.~(1999) have compiled an extensive  catalogue of
parameters for a sample of 61 clusters, including their
distances. These have been obtained by using, as a standard candle,
different theoretical determinations of the HB brightness (see Ferraro
et al.~1999 for details).  Again, the Ferraro et al.~(1999)
observational database, as the Harris~(2003) one, is not photometrically homogeneous.
Carretta et al. ~(2000, 2003) have provided the distance to a small
sample of GGCs by adopting the MS fitting method.  Due to the
impressive care devoted in deriving the subdwarfs parameters such as
metallicity, color, and absolute visual magnitude, as well as cluster
metallicity and reddening, their distance 
measurements appear the most
accurate ones presently available for GGCs. Therefore, despite the
small number of objects involved, the Carretta et al.  measurements
provide an important check of the accuracy and reliability of the
distances presented in this paper.

It is worth remembering that our apparent distance modulus
determinations have been obtained in the HST F555W band. Even if this
photometric band is similar to the Johnson visual band (used in the
other works), it is not exactly the same.  In order to perform a
meaningful comparison, for clusters with [Fe/H] $<$ -0.8, we transformed the F555W apparent distance
modulus estimates into extinction corrected ones, by using our
estimates of E($F439W-F555W$), and the relation presented by Holtzman
et al.\ (1995, Table 12). For comparison purposes, we also transformed
the E($B-V$) values of Harris~(2003) into the
corresponding E($F439W-F555W$) reddenings by calculating the
extinctions coefficents A$_{F555W}$ and A$_{F439W}$ following Holtzman
et al.\ (1995).
 
Fig.~4 shows a comparison of our reddenings and those of Harris
(2003). There is an overall good agreement down
E($F439W-F555W$)$\sim0.75$, though the dispersion of the differences
increases at increasing reddening. For reddenings larger than 0.75,
there seems to be a systematic trend. This is very likely due to
problems in the transformations of Harris' E($B-V$) to the
E($F439W-F555W$).  The transformations of Holtzman et al.\ (1995) from
the E($B-V$) in the Johnson system to the extinctions coefficents in
the WFPC2 flight system are likely less reliable for E($B-V$)$>0.75$.
On the other hand, our reddening measurements, as derived from the
overlap of the object and the template CMDs, could be also more uncertain
for very high, sometimes differential, reddenings.

In  Figs.~5 and 6,  we perform  a comparison  between  our absolute
distance modulus determinations and those of Harris (2003), Ferraro et
al. (1999), and Carretta et al.  (2000, 2003) as a function of cluster
reddening and metallicity, respectively. 
Because of the problems on the transformation of reddenings to the flight systems 
for high extinctions, we did not use clusters with
E($F439W-F555W$)$>0.75$ (open symbols in the figures) in the calculations 
of the mean differences. Similarly, we did not include clusters with [Fe/H]$>$-0.8 (triangles) 
because we had no E($F439W-F555W$) values to perform
the transformation from apparent to absolute distance moduli
(for these clusters the E(B-V) from Harris 2003 catalogue were used to plot the differences in the figure). 

The results of the comparisons in Fig.~5 and 6 can be summarized as
follows:

\begin{itemize} 

\item The mean  difference between  our estimates  and  the Harris's
(2003) ones is of 0.09 magnitudes, our distances being on
average larger. This fact can be easily accounted for when considering
the brightness difference between the theoretical HB luminosity
adopted in the present work and the one adopted by Harris (2003). The
dispersion of the residuals around the mean value is equal to 0.11
magnitudes.
 
\item When comparing our data with those by Ferraro et al. (1999), we
obtain a mean difference of 0.09 mag., with a dispersion of about 0.17
magnitudes. Once again, our distance modulus estimates are larger than
those provided by Ferraro et al.  (1999). This difference is mainly
due to the fact that our theoretical values of the HB luminosity at
the RR Lyrae instability strip are brighter by $\approx0.10$ mag than
those adopted by Ferraro et al.~(1999).  It is not so clear the origin
of the larger dispersion and the apparent dependence of the
differences on metallicity. The two clusters with the highest
disagreement are NGC 6584 and NGC 2808. However, Ferraro et al.'s
distance moduli for these clusters are inconsistent with Harris'
estimates too, with differences of the order of 0.3 mag, Ferraro et al.'s values being
lower. On the other hand, there is a good agreement between ours and
Harris's distances for these two specific objects. In fact, the
difference (our paper--Harris catalogue) is 0.09 for both clusters,
perfectly consistent with the average zero point difference between us
and Harris.  Finally, we note that the disagreement
between Ferraro's and Harris' values is particularly high for
intermediate metallicity clusters.
  
\item The  comparison with the data  by Carretta et  al.~(2000, 2003)
shows  that  the  average  difference  in the  derived  true  distance
estimates is of the order of only -0.015 mag and the  dispersion around the mean
value of the difference is of the order of 0.084 mag.
 
\end{itemize}

\noindent
Our distance modulus zero point appears to be in good agreement with
those provided by Carretta et al.~(2000, 2003).  Even if this
comparison is possible only for a very small number of clusters, this
evidence strongly supports the accuracy and reliability of our
distance estimates.
In addition, the dispersion of the differences is smaller than the difference
rms in the comparisons with Harris~(2003) and Ferraro~(1999) et al.\,
further strengthening the overall accuracy of our distance estimates.
  
\section{Final remarks} 
 
We have employed a large sample of GGC CMDs, obtained and
analized in a fully homogeneous and self-consistent framework, to
estimate the apparent cluster ZAHB luminosity levels as well
as the cluster reddenings.

By using updated stellar evolution models, and in particular new
predictions about the ZAHB luminosity level, we have provided an
estimate of the distances to all clusters. Even if we are aware of
remaining systematic uncertainties which can affect theoretical ZAHB
absolute magnitudes, we are confident that at least our relative
distances are reliable. In addition, we remark that by using the
apparent m$^{ZAHB}_{F555W}$ values listed in table 3) and 4), which
are completely model independent, any interested reader can derive
distance estimates by using the preferred theoretical
framework.

In order to assess the intrinsic accuracy of the present results, we
have performed a comparison between current data and similar
measurements presented by Harris (2003), Ferraro et al. (1999) and
Carretta et al. (2002, 2003). This comparison showed that there are
some problems in the determination of the extinction coefficients in
the WFPC2 flight system from the classical E($B-V$) system for
E($B-V$)$>$0.75, using Holtzman et al. (1995) recipe.

We have also to notice the fine agreement achieved in the comparison 
with the empirical MS-fitting distances by Carretta et al.  (2000,
2003). This lends strong support to both our relative and absolute
ZAHB distance scale. Accurate empirical analysis 
like the Carretta et al. ones, extended to a larger sample of objects,
are needed in order to definitely  confirm the reliability of
our ZAHB absolute distances.

\begin{acknowledgements} 
ARB recognizes  the  support of   the Istituto Nazionale di Astrofisica({\it INAF}).
GP and SC recognize partial support from the Ministero dell'Istruzione, Universit\`a 
e Ricerca ({\it MIUR}, PRIN2002, PRIN2003), and from the Agenzia Spaziale Italiana 
({\it ASI}). 
\end{acknowledgements}

\end{document}